\documentstyle[aps]{revtex}
\begin{document}
\title{
Speed Meter as a Quantum Nondemolition Measuring Device for Force
}
\author{F.\ Ya.\ Khalili$^{(1)}$ and Yu. Levin$^{(2)}$ 
} 
\address{$^{(1)}$Physics Faculty, Moscow State University, Moscow Russia}
\address{$^{(2)}$Theoretical Astrophysics, California Institute of
Technology, Pasadena, California 91125}
\date{\today}
\maketitle
\begin{abstract}
Braginsky has proposed a {\it speed meter} (a speed or momentum measuring
device), consisting of a small Fabry-Perot cavity 
rigidly attached to a freely moving test mass.  This paper devises an
optical readout strategy which enables the
meter, when monitoring a classical force via speed changes, to beat 
the standard quantum limit---at least in principle.  

\end{abstract}
\pacs{PACS numbers: XXXXXXXXX}

\narrowtext

\section{Introduction}

A laser interferometer gravitational wave detector is, in essence, a
device for monitoring a classical force (the gravitational wave) that
acts on freely moving test masses (the interferometer's suspended
mirrors).  ``Advanced'' detectors, expected to operate in the LIGO/VIRGO
interferometric network\cite{ligovirgo}  in the middle or later part of 
the next decade,
will be constrained by the {\it standard quantum limit} (SQL) for force
measurements\cite{braginsky_khalili,300yrs},
\begin{equation}
\Delta F_{\rm SQL} = \sqrt{\hbar m / \tau^3}\;.
\label{eq:sql}
\end{equation}
Here $\hbar$ is Planck's constant, $m$ is the mass of the test body on
which the force acts, and $\tau$ is the duration of the force.
 
It is well known that the SQL is not an absolute barrier to further
sensitivity improvements\cite{braginsky_khalili}.  With cleverness, one
can devise so-called {\it quantum nondemolition} (or QND) measurement
schemes, which beat the SQL.  Although fairly practical QND techniques have
actually been devised for  
resonant-mass gravitational-wave detectors\cite{bvt}, no practical QND
technique yet exists for the interferometric gravitational-wave
detectors on which LIGO/VIRGO is based.  The effort to devise such a
technique is of great importance for the long-term future of the LIGO/VIRGO
network. 

Although a practical QND technique for such detectors is not yet known,
several idealized techniques have been formulated\cite{qndideas} and are
playing helpful roles in the search for a practical technique.  Most of
these idealized techniques are based on optical measurements of a test mass's
position.  One, however, is based on measurements of the mass's speed
or momentum.  This ``speed meter'' has been devised in initial,
conceptual form by Braginsky\cite{braginsky}, and he has argued
that it should be capable of beating the SQL.  

The purpose of this paper is to demonstrate that, when coupled to a
specific optical readout scheme that we have devised, Braginsky's speed
meter does, indeed, beat the SQL, at least in principle. 

\section{The Basic Idea of the Speed Meter}

The fundamental idea underlying Braginsky's speed meter is to attach a
small, rigid Fabry-Perot cavity to the test mass, whose speed is to be
measured.  The cavity's two mirrors are to have identical
transmissivities and negligible losses, in the idealized variant we
shall analyze.  This means that, when the cavity is at rest and is
excited by laser light that is pricisely in resonance with one of the
cavity's modes, the light passes straight through the cavity without
reflection and emerges from the other side unchanged.  When the cavity
starts moving, by contrast, it sees the incoming light Doppler shifted;
and, as a result, the light emerging from the other side gets phase
shifted by an amount
\begin{equation}
\Delta\phi = {\omega_o v \tau \over c}\;,
\label{eq:phaseshift}
\end{equation}
where $v$ is the cavity's speed, $\omega_o$ is its
eigenfrequency,  and $\tau$ is its ringdown time.  Here
and throughout we assume that $v\ll c/(\omega_o\tau)$.

By measuring the phase of the emerging light, one can infer the speed of
the cavity without learning anything about its position.  This absence
of information about position implies
(Braginsky has argued) that such a device should be able to
evade any back-action force of the measurement on the cavity's
velocity (or momentum), and therefore should be capable of beating the SQL.  

\section{Our Readout Scheme for the Speed Meter, and an Analysis of its
Performance}

In this section we shall exhibit an optical readout scheme for such a
speed meter which does, indeed, enable it to 
 beat the SQL.  Our readout scheme is sketched
in Fig.\ \ref{fig:readout}, whose details will become more clear in what
follows.

Let
\begin{equation}
\psi_{in}=Ae^{-i\omega_{0}(t- x/ c)}+\int_{0}^{\infty} d\omega\sqrt{\hbar\omega\over sc}
 a(\omega)e^{-i\omega(t-x/ c)} 
\end{equation}
be the incoming field  on the left side of the cavity. The first term on the right hand side 
of (3) represents classical pumping, and the second term shows quantum fluctuations of the incoming field;
$x$  is the  position of the cavity and acquires time dependence when the cavity is moving;
$s$ is the area of the beam. Also   $a(\omega)$, $b(\omega)$, $c(\omega)$ and $d(\omega)$
represent annihilation 
operators for four modes as in Fig.1, and   $t(\omega)$ and $r(\omega)$ represent frequency-dependent
tranmission and reflection coefficients of the cavity respectively. In our set-up $t(\omega_{0})=1$,
$r(\omega_{0})=0$.

 Any motion of the cavity induces a time-dependence of $x$ in (3),
and hence with respect to the cavity the classical pump acquires frequencies different from $\omega_{0}$.
The resulting effect 
of the cavity motion is to scatter the classical part of the incoming wave into modes which would
otherwise carry only vacuum fluctuations. A simple calculation produces the following relations:

\begin{eqnarray}
 c(\omega)&=&t(\omega) a(\omega)+r(\omega) b(\omega)\nonumber\\
          & &+iA{\omega_{0}\over c} \sqrt{sc\over\hbar\omega}
      (t(\omega)-1)\tilde{X}(\omega-\omega_{0}),\label{line1}\\
 d(\omega)&=&r(\omega)a(\omega)+t(\omega)b(\omega)\nonumber\\
          & &+iA{\omega_{0}\over c} \sqrt{sc\over\hbar\omega}             
      r(\omega)\tilde{X}(\omega-\omega_{0}),\label{line2}
\end{eqnarray}
where $\tilde{X}(\Omega)$ is defined by 

\begin{equation}
x(t)=\int_{-\infty}^{\infty} \tilde{X}(\Omega) e^{-i\Omega t} d\Omega
\end{equation}

We assume that the cavity is pushed by an external signal force $F_{\rm s}(t)$
(due, e.g., to a gravitational wave). Then it's position obeys the 
free-mass equation of motion
\begin{equation}
F=F_{\rm s}+F_{\rm fl}=m\ddot{x},
\end{equation}
where $F_{\rm fl}$ stands
for the random force produced by quantum fluctuations of the light. This force, as
evaluated using momentum concervation, has the following  Fourier transform           

\begin{eqnarray}
 \tilde{F}_{\rm fl}(\Omega)&=&\sqrt{W\hbar \omega_{0}\over 2\pi c^{2}}\left\{
                               \left[ 1-t\left(\omega_{0}+\Omega\right)\right]
                               \left[a\left(\omega_{0}+\Omega\right)+
                               a^{\dagger}\left(\omega_{0}-\Omega\right)\right]\right.\nonumber\\
                           & &\left.-r\left(\omega_{0}+\Omega\right)
                              \left[b\left(\omega_{0}+\Omega\right)+
                                b^{\dagger}\left(\omega_{0}-\Omega\right)\right] \right\}  \label{line1}
\end{eqnarray}
for $\Omega\ll\omega_{0}$, where $W={scA^{2}/ 2\pi}$ is the power of the incoming wave. Formula 
(8) can be simplified by noting that for the Fabry-Perot cavity $t+r=1$, so

\begin{eqnarray}
\tilde{F}_{\rm fl}(\Omega)&=&\sqrt{W\hbar\omega_{0}\over 2\pi c^{2}}r(\omega_{0}+\Omega)[
                             a(\omega_{0}+\Omega)-b(\omega_{0}+\Omega)\nonumber\\
                          & &+
                             a^{\dagger}(\omega_{0}-\Omega)-b^{\dagger}(\omega_{0}-\Omega)] .
                            \label{line2}
\end{eqnarray}
It is clear from Eq.(9) that $f(\omega)=[a(\omega)-b(\omega)]/ \sqrt{2}$ is the only combination
of the incoming modes which  appears, multiplied by $x$, in the interaction part of the 
 Hamiltonian .  Therefore, all information about the motion of the cavity should be recorded
in $f$ and $f^{\dagger}$. 
The obvious suitable choice of readout is 
\begin{equation}
 e(\omega)={c(\omega)-d(\omega)\over\sqrt{2}},
\end{equation}
since putting (6), (7), (9) into (10), we can express it as solely a function of $f$ and the signal
force:
\begin{eqnarray}
 & &e(\omega_{0}+\Omega)=[t(\omega_{0}+\Omega)-r(\omega_{0}+\Omega)]f(\omega_{0}+\Omega)+\nonumber\\
 & &                   2i{W\omega_{0}\over m\Omega^{2} c^{2}}r(\omega_{0}+\Omega)^{2}\left[f(\omega_{0}+\Omega)+
                         f^{\dagger}(\omega_{0}-\Omega)\right]\label{line1}\\
& &+\sqrt{2}i{\omega_{0}\over c}\sqrt{2\pi W\over\hbar\omega_{0}}r(\omega_{0}+\Omega) 
                       {F_{\rm s}(\Omega)\over m\Omega^{2}} .\nonumber
\end{eqnarray}
It is attractive to read out $e(\omega)$ using homodyne detection as sketched in Fig. 1. The measurement
output then is the homodyne quadrature
\begin{equation}
y(\Omega)=e(\omega_{0}+\Omega)e^{i\psi(\Omega)}+e^{\dagger}(\omega_{0}-\Omega)e^{-i\psi(\Omega)}, 
\end{equation}
where $\psi(\Omega)$ is a phase factor that we shall fix below so as to minimize the noise (see also \cite{vyatchanin}).
Then the quantum noise spectral density in this measured quantity, as computed from the formula
$\langle y(\Omega) y(\Omega^{\prime})\rangle =S_{\rm y}(\Omega)\delta (\Omega+\Omega^{\prime})$, is
\begin{eqnarray}
S_{\rm y}(\Omega)&=&  
                 2\alpha^{2}(\Omega)\left\{1-\cos\left[2\psi\left(\Omega\right)\right]-\right.\nonumber\\
                 & &\left.2\alpha\left(\Omega\right)\sin\left[2\psi\left(\Omega\right)
                                         \right]\right\}+1\label{line1}            
\end{eqnarray}
where $\alpha(\Omega)=(2W\omega_{0}/ m\Omega^{2}c^{2})|r(\omega_{0}+\Omega)|^{2}$.
This noise is minimized for
\begin{equation}
\tan2\psi_{\rm min}(\Omega)={1\over \alpha(\Omega)}.
\end{equation}
Putting in explicitly $r(\Omega)=i\Omega\tau_{\rm ringdown}/( 1+i\Omega\tau_{\rm ringdown})
$ we see that for large power $W\gg mc^2/\omega_{0}\tau_{\rm ringdown}^{2}$ the minimum noise is 
\begin{equation}
S_{\rm y}={1\over 4\alpha(\Omega)^{2}}={{m^2}{\Omega^4}{c^4}\over 16 {W^2}{\omega_{0}^2}
          |r(\omega_{0}+\Omega)|^2}.
\end{equation}
 Here $\tau_{\rm ringdown}$ is the e-folding time 
for resonant light to escape from the cavity. Now suppose that the form of the signal 
, $F_{\rm s}(\Omega)$, is known and we use an optimal filter to search in the output 
$y(\Omega)$ to see whether the signal is actually present. The  signal to noise ratio
for this search is given by  
\begin{equation}
{S\over N}={1\over 2\pi}\int_{\Omega_{1}}^{\Omega_{2}}{16\pi W\omega_{0}|r(\omega_{0}+\Omega)|^{2}
\over mc^{2}\Omega^{2}} {|F_{\rm s}(\Omega)|^{2}\over\hbar m\Omega^{2}}d\Omega .
\end{equation}
For a narrow-band measurement ($\tau_{\rm meas}\gg\tau_{\rm ringdown}$)  $\psi(\Omega)$ is a constant
($\psi\rightarrow mc^{2}/ 4W\omega_{0}\tau_{\rm ringdown}^{2}$) which makes homodyne detection technically
easier (see Fig.1), and we have
\begin{equation}
{S\over N}={W\over W_{\rm SQL}} \left({S\over N}\right)_{\rm SQL} ,
\end{equation}
where
\begin{equation}
\left({S\over N}\right)_{\rm SQL}={1\over \hbar m}\int{|F_{\rm s}|^{2}\over \Omega^{2}}d\Omega
\end{equation}
and
\begin{equation}
W_{\rm SQL}={mc^{2}\over 16\pi\omega_{0}\tau_{\rm ringdown}^{2}} .
\end{equation}
Thus the minimum detectible force  may  be lower than the Standard Quantum Limit by a factor of 
$\sqrt{W_{\rm SQL}/W}$:
\begin{equation}
F_{\rm min}=\sqrt{W_{\rm SQL}\over W}\Delta F_{\rm SQL}.
\end{equation}
The minimum detectible force is proportional to $1/\sqrt{W}$. 
Eq. (20) obviously holds for a broad-band signal as well, but the 
expression for $W_{\rm SQL}$ is different and the homodyne phase 
acquires frequency dependence, which makes the detection very difficult.

\section{Conclusion}

We have shown that, for the readout scheme of Fig.\ \ref{fig:readout},
the minimum measureable force $F_{\rm min}$ is given by
\begin{equation}
F_{\rm min} = \sqrt{W_{\rm SQL}\over W} \Delta F_{\rm SQL}\;,
\label{eq:Fmin}
\end{equation}
where $\Delta F_{\rm SQL}$ is the standard quantum limit [Eq.\
(\ref{eq:sql})], $W$ is the laser power, and $W_{\rm SQL}$ is the
minimum laser power required for beating the SQL:
\begin{eqnarray}
W_{\rm SQL} &=& {mc^2\over 16\pi\omega_o\tau^2} \nonumber
\\
&=& 5\times 10^4 \hbox{Watt}
{m\over 10{\rm kg}} {4\times 10^{15} {\rm s}^{-1}\over \omega_o} \left({0.01{\rm
s}\over \tau}\right)^2 
\label{eq:Wsql}
\end{eqnarray}

Although it is not outrageous to imagine achieving the laser powers $W>W_{\rm
SQL}$ at which Eq.\ (\ref{eq:Fmin}) reports a beating of the SQL, there
are many serious practical obstacles to implementing such a speed meter
in a real interferometric gravitational-wave detector.  Nevertheless,
this speed meter might contain the conceptual seeds from which will grow
a practical QND scheme for the LIGO/VIRGO network.

\section{Acknowledgments}
We thank Vladimir Braginsky for proposing this research and for technical
advice during its execution, and Kip Thorne for advice about the prose.
Y.L. also thanks Kip Thorne and Ron Drever
for useful discussion and advice.
This research was supported in part by NSF Grants XXXX [the
Braginsky/Whitcomb grant] and PHY-9424337, and by XXXXX.

\newpage

\begin{figure}
\caption[]{
The speed meter and its optical readout.
}
\label{fig:readout}
\end{figure}

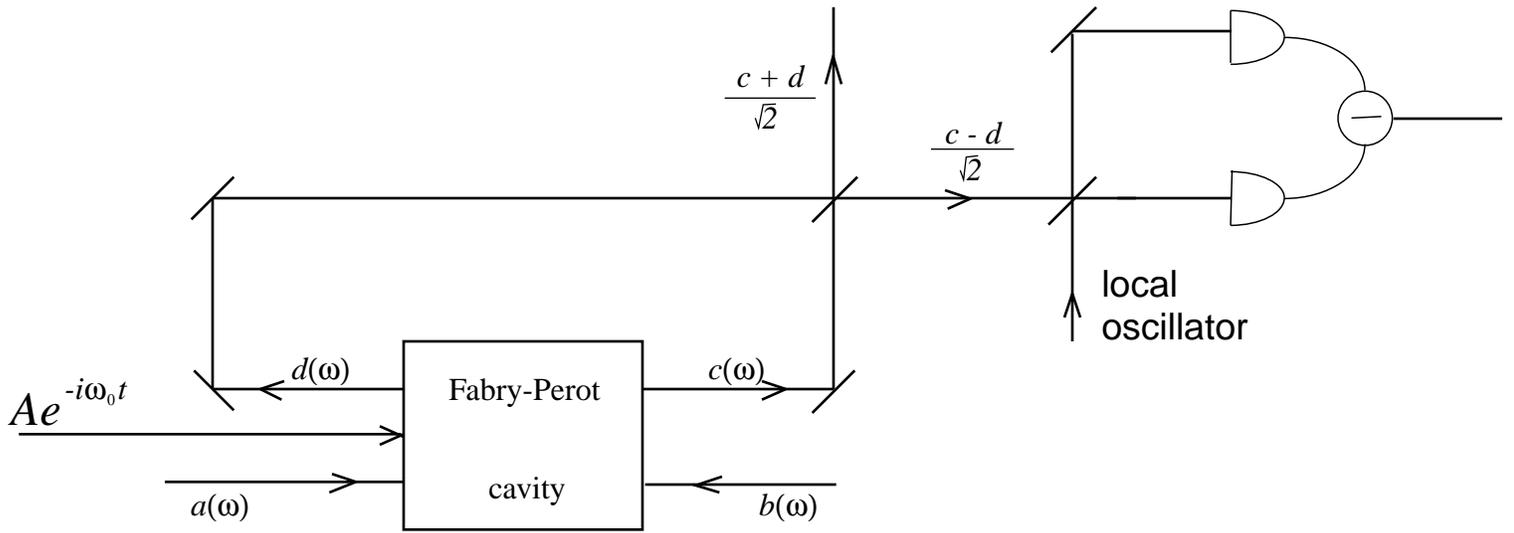
\end{document}